\documentclass[]{scrartcl}
\usepackage{graphics}
\usepackage[english]{babel}
\usepackage{amsmath} 
\usepackage{amsfonts}
\usepackage{latexsym}
\usepackage{graphicx}
\usepackage{graphics}
\usepackage{float}

\usepackage{amssymb}
\usepackage{authblk}
\usepackage{multirow}
\usepackage{makeidx}
\usepackage{dblfloatfix}
\usepackage[section]{placeins}

\PassOptionsToPackage{hyphens}{url}
\usepackage{hyperref}
\usepackage{url}
\begin{document}

\title{A Test of General Relativity Using the LARES and LAGEOS Satellites and a GRACE Earth's Gravity Model}

\subtitle{Measurement of Earth's Dragging of Inertial Frames}

\author[1,2]{Ignazio Ciufolini\thanks{ignazio.ciufolini@unisalento.it}}
\author[2,3]{Antonio Paolozzi}
\author[4]{Erricos C. Pavlis}
\author[5]{Rolf Koenig}
\author[6]{John Ries}
\author[7]{Vahe Gurzadyan}
\author[8]{Richard Matzner}
\author[9]{ Roger Penrose}
\author[10]{Giampiero Sindoni}
\author[2,3]{Claudio Paris}
\author[7]{Harutyun Khachatryan}
\author[7]{ Sergey Mirzoyan}

\affil[1]{\footnotesize Dip. Ingegneria dell'Innovazione, Universit\`a del Salento, Lecce, Italy}
\affil[2]{Museo della fisica e Centro studi e ricerche Enrico Fermi, Rome, Italy}
\affil[3]{Scuola di Ingegneria Aerospaziale, Sapienza Universit\`a di Roma, Italy}
\affil[4]{Joint Center for Earth Systems Technology (JCET), University of Maryland, Baltimore County, USA}
\affil[5]{Helmholtz Centre Potsdam, GFZ German Research Centre for Geosciences, Potsdam, Germany}
\affil[6]{Center for Space Research, University of Texas at Austin, Austin, USA}
\affil[7]{Center for Cosmology and Astrophysics, Alikhanian National Laboratory and Yerevan State University, Yerevan, Armenia}
\affil[8]{Theory Center, University of Texas at Austin, Austin, USA}
\affil[9]{Mathematical Institute, University of Oxford, Oxford, UK}
\affil[10]{DIAEE, Sapienza Universit\`a di Roma, Rome, Italy}
\renewcommand\Authands{ and }

\maketitle

\begin{abstract}
We present a test of General Relativity, the measurement of the Earth's dragging of inertial frames. Our result is obtained using about 3.5 years of laser-ranged observations of the LARES, LAGEOS and LAGEOS 2 laser-ranged satellites together with the Earth's gravity field model GGM05S produced by the space geodesy mission GRACE. We measure $\mu = (0.994 \pm 0.002) \pm 0.05$, where $\mu$ is the Earth's dragging of inertial frames normalized to its General Relativity value, 0.002 is the 1-sigma formal error and 0.05 is the estimated systematic error mainly due to the uncertainties in the Earth's gravity model GGM05S. Our result is in agreement with the prediction of General Relativity.
\end{abstract}

\section{Introduction}
\label{intro}
About one hundred years ago Albert Einstein completed
the publication of a series of fundamental papers describing the gravitational theory known as General Relativity
(GR) \cite{ein1,ein2,ein3,ein4}. Since then Einstein's gravitational theory
has had experimental and theoretical triumphs,
including the prediction and observation
of the expansion of the universe, of black holes,
gravitational lensing and gravitational waves \cite{nov,wei,nov,mtw,ciuw,tur,wil,ligo}. GR has today a number of
practical applications to our everyday life \cite{kop} including its corrections that enable the Global Navigation Satellite
System to reach accuracies at the level of a few decimeters
\cite{ash}.

Nevertheless, GR has not been reconciled with the other
fundamental theory of modern physics: Quantum Mechanics. Further, Einstein's
gravitational theory predicts the occurrence of spacetime
singularities where every known physical theory ceases to
be valid, the spacetime curvature diverges and time ends \cite{pen}.
In 1998 observations of distant supernovae of type Ia
implied the quite surprising result
that the universe has an accelerated expansion \cite{sup,sup2}.
An explanation for this mysterious result can be found in
the cosmological constant introduced by Einstein to avoid
a dynamical universe and later, in 1931, abandoned by
Einstein himself. However, the cosmological constant
corresponds to vacuum energy and quantum field theory
predicts that the vacuum energy should have a value
approximately $10^{122}$ times larger than the {\it dark energy} \cite{darke,darke2}
density that is observed in the universe.
To explain the accelerated expansion of the universe, dark energy
should compose more than 70\% of our universe, but
its real nature is unknown. Other explanations include
a time dependent vacuum energy with the exotic
name of quintessence, and modifications of GR such as the
so-called f(R) theories. Therefore, in spite of its
experimental triumphs, Einstein's gravitational theory continues to
need further accurate tests at all scales from solar system
tests to astrophysical and cosmological observations.

Successful tests  \cite{ciuw,tur,wil} of effects and phenomena predicted by GR include the well known perihelion precession of Mercury (and in general the periastron advance of
an orbiting body), the equivalence principle and the time-dilation of clocks in a gravitational field, the deflection and time-delay of electromagnetic waves by a mass, the dynamics of the Moon, accurately measured by Lunar Laser Ranging and of binary pulsars \cite{tay,ocon,ocon2}, gravitational lensing and other relevant astrophysical observations. Gravitational waves have been indirectly observed at the level predicted by GR from the rate of change of the orbital period of the binary pulsar PSR B1913+16 \cite{tay}.
Recently the two LIGO advanced detectors (Caltech and MIT) have directly detected the gravitational waves from the inspiral and merger of a pair of black holes \cite{ligo} marking the beginning of the gravitational-wave astronomy.

\section{Dragging of Inertial Frames}
\label{sec:1}

Among the intriguing phenomena predicted by GR, and
so far only tested with approximately 10\% accuracy, is the "dragging of inertial
frames", or "frame-dragging" as Einstein named it in 1913
\cite{ciunat07}. Frame-dragging has relevant astrophysical applications
to the dynamics of matter falling into rotating black holes
and of jets in active galactic nuclei and quasars \cite{tho}.

A test-gyroscope is a small current of mass in a loop and
may be realized using a sufficiently small spinning top. In
GR a gyroscope determines the axes of local nonrotating
inertial frames. In such frames the equivalence principle
holds so that the gravitational field is locally unobservable
and all the laws of physics are the laws of Special
Relativity theory. However in GR a gyroscope has a potential
behaviour different from that in classical Galilei-Newton mechanics.
In classical mechanics, a torque-free gyroscope is predicted to
always point towards the same distant ``fixed'' stars. In contrast,
in GR, a gyroscope is dragged by mass currents, such
as the spinning Earth, and therefore its orientation can change
with respect to the distant ``fixed'' stars. If we were to rotate
with respect to the gyroscope, we would feel centrifugal
forces, even though we may not rotate at all with respect
to distant ``fixed'' stars \cite{ciuw}.

Frame-dragging of a gyroscope is formally similar to the change of orientation of a magnetic dipole by a magnetic field generated by an electric current in electrodynamics \cite{tho}. In GR, a current of mass generates an additional contribution to the gravitational field, called “gravitomagnetic field” because of its formal analogy with electrodynamics. The gravitomagnetic field then exerts
a torque on a gyroscope in the same way a magnetic
field torques a magnetic needle in electrodynamics.

In 1918, Lense and Thirring \cite{len} published the equations of the frame-dragging perturbations of the orbital elements of a satellite in the weak gravitational field of a slowly rotating body. The rate of change of the nodal longitude of the satellite, known as Lense-Thirring effect, is given by $\dot {\bf \Omega} \, = \,{2 {\bf J} \over a^3 \, (1 - e^2)^{3/2}}$, where ${\bf \Omega}$ is the nodal longitude of the satellite,       $a$ its semimajor axis, $e$ its orbital eccentricity and ${\bf J}$ is the angular momentum of the rotating body. We recall that the node, ascending or descending, of a satellite is defined as the intersection of its orbit with the equatorial plane of the central body, in our case the Earth \cite{kau}.

Frame-dragging was observed \cite{ciusci} in 1997-1998 by using the LAGEOS (LAser GEOdynamics Satellite) and LAGEOS 2 laser-ranged satellites \cite{lag} and measured with approximately 10\% accuracy \cite{ciunat,ciupavper,ciuwhe,ciuepjp} in 2004-2010, using LAGEOS, LAGEOS 2 and the Earth's gravity field determinations by the space geodesy mission GRACE \cite{grace1,grace2}. In 2011 the dedicated space mission Gravity Probe B, launched in 2004 by NASA, reported also a test of frame-dragging with approximately 20\% accuracy \cite{GPB}.

LAGEOS was launched in 1976 by NASA, and LAGEOS 2 in 1992 by ASI and NASA \cite{lag}. They are two almost identical passive satellites covered with 426 corner cube reflectors to reflect back the laser pulses emitted by the stations of the Satellite Laser Ranging (SLR) network \cite{ilrs}. SLR allows measurement of the position of the LAGEOS satellite with an accuracy that can reach a few millimetres over a range of about 6000 km. The twin GRACE (Gravity Recovery and Climate Experiment) satellites were  launched in 2002 by NASA and DLR (the German Aerospace Center). They are 200 - 250
km apart, in a near-polar orbit at an altitude of about 480 km. The GRACE space mission has allowed extremely
accurate determinations of the Earth's gravitational field and its temporal variations. For the main characteristics and orbital parameters of LARES, LAGEOS, LAGEOS 2 and GRACE see Table 1.
\begin{table*}
	\begin{center}
		\caption{Main characteristics and orbital parameters of the satellites used in the LARES  experiment.}
		\label{tab:2}
		\begin{tabular}{lllll}
			\hline
			& LARES & LAGEOS & LAGEOS 2 & GRACE \\
			\noalign{\smallskip}\hline\noalign{\smallskip}
			Semimajor axis [km]& 7821  & 12270  & 12163  & 6856  \\
			\noalign{\smallskip}\hline\noalign{\smallskip}
			Eccentricity & 0.0008 & 0.0045 & 0.0135 & 0.005 \\
			\noalign{\smallskip}\hline\noalign{\smallskip}
			Inclination & $69.5^\circ$ & $109.84^\circ$ & $52.64^\circ$  & $89^\circ$ \\
			\noalign{\smallskip}\hline\noalign{\smallskip}
			Launch date & 13 Feb, 2012 & 4 May, 1976 & 22 Oct, 1992 & 17 Mar, 2002 \\
			\noalign{\smallskip}\hline\noalign{\smallskip}
			Mass [kg]& 386.8  &406.965  & 405.38  & 432  \\
			\noalign{\smallskip}\hline\noalign{\smallskip}
			Number of CCRs & 92 & 426 & 426& 4  \\
			\noalign{\smallskip}\hline\noalign{\smallskip}
			Diametre [cm] & 36.4  & 60  & 60  &  \\
			\hline
		\end{tabular}
	\end{center}
\end{table*}
The test of frame-dragging with the LAGEOS satellites was obtained by using the two observables quantities given by the two nodal rates of LAGEOS and LAGEOS 2 for the two main unknowns: the frame-dragging effect and the uncertainty in the Earth's quadrupole moment, $J_2$ \cite{ciuncc}. If the Earth's gravitational potential is expanded in spherical harmonics, the even zonal harmonics are those of even degree and zero order. They represent the deviations from spherical symmetry of the gravitational potential of a body which are axially symmetric and which are also symmetric with respect to the equatorial plane of the body. The main secular drifts of the nodal longitude of a satellite are due to the Earth's even zonal harmonics. In particular, the largest node shift is by far due to the even zonal of degree two, $J_2$, i.e. the Earth's quadrupole moment \cite{kau}. To measure frame-dragging we either need to perfectly determine the Earth's even zonal harmonics or devise a method to neutralize the propagation of their uncertainties in our measurement.


\section{LARES}
\label{sec:2}
LARES is a satellite of the Italian Space Agency (ASI) launched by the European Space Agency with the new launch vehicle VEGA (ESA-ASI-ELV-AVIO). It is a passive, spherical laser-ranged satellite (see Table 1). The LARES satellite was designed to approach as closely as possible an ideal test particle \cite{pao1}. This goal was mainly achieved by adopting the following design requirements: (i) minimize the surface-to-mass ratio, (ii) reduce the number of parts, (iii) avoid any protruding component, (iv) use a non magnetic material and (v) avoid the painting of the satellite surface. The first requirement was implemented by using a tungsten alloy \cite{pao2}, the most dense material on Earth with an acceptable cost and good manufacturing characteristics. With a diameter of 36.4 cm and a total mass of 386.8 kg, the final mean density of the satellite is 15317 $kg/m^3$ which makes LARES the known orbiting object in the solar system with the highest mean density and the satellite with the lowest surface-to-mass ratio. The second requirement was achieved by building the satellite body out of one single piece of tungsten alloy, thus reducing thermal contact conductance and consequently the onset of thermal gradients. Temperature differences on the surface of the LAGEOS satellites produce in fact a tiny but not negligible perturbation: the thermal thrust \cite{pao3}. To comply with the third requirement, the satellite interface with the separation system was limited only to four hemispherical cavities machined on the equator of the satellite. The fourth and fifth requirements were simply fulfilled by choosing a non magnetic tungsten alloy, although with slightly lower density than a magnetic tungsten alloy, with a proper surface treatment and with no painting \cite{pao5}.

\section{Test of frame-dragging using LARES and the two LAGEOS satellites}

The basic idea of the LARES space mission is to couple its orbital data with those of the two LAGEOS satellites in order to have three observable quantities provided by the nodal rates of the three satellites \cite{ciuwhe2}. The three observables can then be used to determine the three unknowns: frame-dragging and the two uncertainties in the two lowest degree even zonal harmonics,  $J_2$, and $J_4$ (i.e. the spherical harmonics of degree
2 and 4 and order 0). In such a way the two largest sources
of uncertainty in the nodal drift are eliminated, providing
an accurate measurement of frame-dragging within our systematic uncertainty of a few percent.

Here we report on our orbital analysis of the laser-ranging data of the LARES, LAGEOS and LAGEOS 2 satellites from 26 February 2012 until 6 September 2015 using
a prominent state-of-the-art Earth's gravity field model, the GGM05S \cite{ggm}. GGM05S is an Earth's gravity model released in 2013, based on approximately 10 years of GRACE data. It describes the Earth's spherical harmonics up to degree 180. The laser-ranging data of LARES, LAGEOS and LAGEOS 2 were collected from more than 30 ILRS stations all over the world (see Figure 1). We processed approximately one million normal points of LARES, LAGEOS and LAGEOS 2, corresponding to about 100 millions of laser ranging observations. The laser-ranging normal points were processed using NASA's orbital analysis and data reduction software GEODYN II \cite{geo}, including the Earth's gravity model GGM05S, Earth's tides,
solar radiation pressure, Earth's albedo, thermal thrust, Lunar, solar and planetary
perturbations and Earth's rotation from Global Navigation Satellite System (GNSS) and Very Long Baseline Interferometry (VLBI).

The orbital residuals of a satellite are obtained by subtracting the observed orbital elements of the satellite with the computed ones. They provide a measurement of the orbital perturbations that, in the data reduction, are not included (un-modelled) or are modelled with some errors (mis-modelled) \cite{ciusci}. In particular, the  residuals of the satellite's node are due to the errors in the Earth's even zonal harmonics and to the Lense-Thirring effect which we have not included in GEODYN II's modelling. The Lense-Thirring nodal shift, theoretically predicted by General Relativity, is about 30.7 milliarcsec/yr on LAGEOS, about 31.5 milliarcsec/yr on LAGEOS 2 and about 118.4 milliarcsec/yr on LARES, the latter corresponding at the altitude of LARES to about 4.5m/yr.

\begin{figure}
	\centering
	\includegraphics[width=0.70\textwidth]{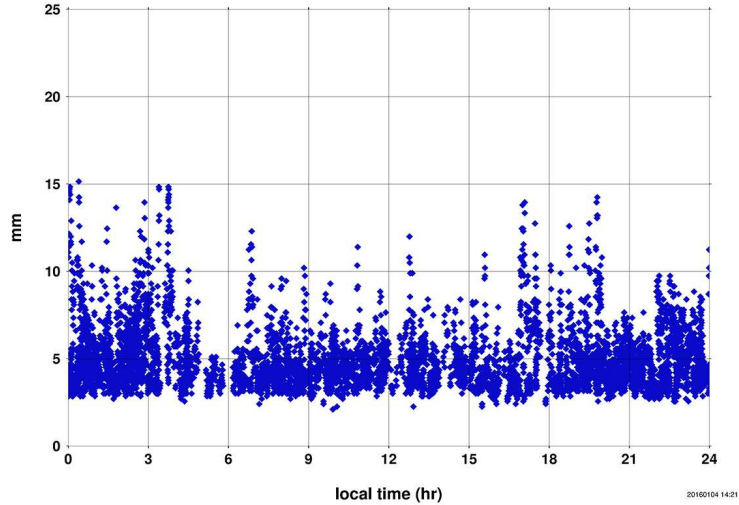}
	\caption{Root Mean Square (RMS) of the LARES normal points obtained from the laser-ranging observations of the Graz station of the ILRS during 2015. The average RMS of the LARES normal points is 4.83 millimeters (Courtesy of the ILRS \cite{ilrs}). }
	\label{fig:1}       
\end{figure}

\begin{figure}
	\centering
	\includegraphics[width=0.7\textwidth]{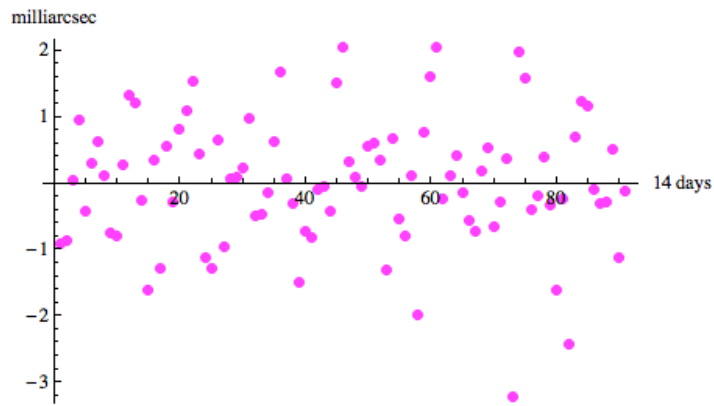}
	\caption{Combined residuals of LARES, LAGEOS and LAGEOS 2, over about 3.5 years of orbital observations, after the removal of six tidal signals and a constant trend.}
	\label{fig:2}       
\end{figure}
Using the three observables provided by the three nodal rates of LAGEOS, LAGEOS 2 and LARES, we were able to eliminate not only the uncertainties in their nodal rates
due to the errors in the even zonal harmonics $J_2$ and $J_4$
of the GGM05S model but also the uncertainties in their nodal rates
due to the long and medium period tides contributing
to the harmonics $J_2$ and $J_4$.

We fitted for the six largest tidal signals of
LAGEOS, LAGEOS 2 and LARES, and for a secular trend, which produced:

\begin{equation}
\mu = (0.994 \pm 0.002) \pm 0.05
\end{equation}

Here $\mu$ = 1 is the value of frame-dragging normalized
to its GR value, 0.002 is the formal 1-sigma error
(the postfit residuals of Figure 2 show a normal--Gaussian--
distribution to good approximation) and 0.05 is the estimated systematic error due
to the uncertainties in
the Earth's gravity field model GGM05S and to the other error souces. We discuss systematic errors below.

In figure 3, we display the least squares secular trend fit of the combined
residuals of LAGEOS, LAGEOS 2 and LARES prior to fitting for the tides.
In contrast, in Figure 4 we show
the secular trend obtained when including the six known
periodical terms corresponding
to the largest tidal signals observed on the satellite's
nodes. The fit is obviously much tighter. These tidal signals
were identified both by a Fourier analysis of the observed
residuals and by analytical computations of the main tidal
perturbations of the nodes of the satellites.
Some of the signals observed in the nodal residuals correspond
to the perturbations due to the main nongravitational perturbations.

\begin{figure}[!h]
	\centering
	\includegraphics[width=0.7\textwidth]{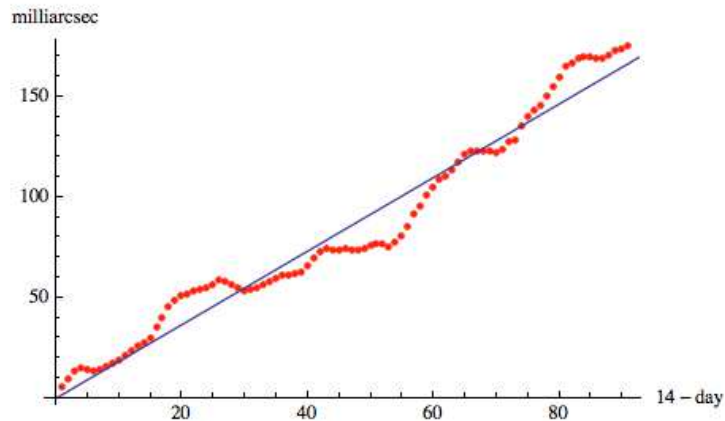}
	\caption{Fit of the combined orbital residuals of LARES, LAGEOS and LAGEOS 2 with a linear regression only.}
	\label{fig:3}       
\end{figure}

\begin{figure}[!h]
	\centering
	\includegraphics[width=0.7\textwidth]{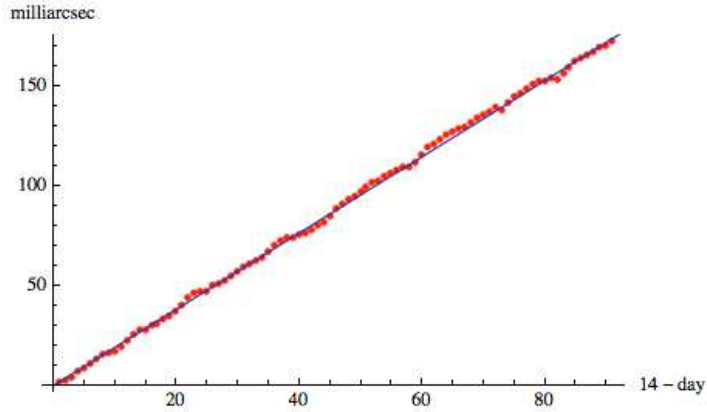}
	\caption{Fit of the combined orbital residuals of LARES, LAGEOS and LAGEOS 2 with a linear regression plus six periodical terms corresponding to six main tidal perturbations observed in the orbital residuals.}
	\label{fig:4}       
\end{figure}

The systematic errors in our measurement of frame-dragging with LARES, LAGEOS and LAGEOS 2 are mainly due to the errors in the even zonal harmonics of GGM05S, used in our orbital fits with GEODYN II, with degree strictly larger than four. To evaluate these systematic errors, we tripled the published calibrated errors
(i.e. including both the statistical and the systematic errors) of each even zonal coefficent of GGM05S (to multiply by a factor two or three is a standard technique in space geodesy to place an upper bound to the real error in the Earth's spherical harmonics) and then propagated these tripled errors into the nodes of LARES, LAGEOS and LAGEOS 2. We then found a systematic error of about 4\% in our measurement of frame-dragging due to the Earth's even zonals.

Other smaller systematic errors are due to those long and medium period tides and non-gravitational perturbations either mis-modelled, or un-modelled. However, in our analysis we included the main tidal and non-gravitational perturbations, such as the direct radiation pressure from the Sun and the Earth, i.e. the albedo. Furthermore, the systematic errors due to the un-modelled or mis-modelled tidal and non-gravitational perturbations are periodical and their residual effect is quite small as clearly shown in the Fourier analysis of the post-fit orbital residuals shown in Figure 2. Previous error analyses \cite{ciu89,rie0,rub,pet,luc,ciupavper,rie1,rie2,rie3,gurc,ciu13} have confirmed that the systematic error due to tides, non-gravitatioanl perturbations and other error sources is at the level of approximately 3\% and therefore the total Root Sum Squared (RSS) systematic error, including the systematic error due to the Earth's even zonals, is approximately at the level of 5\% if the LARES, LAGEOS and LAGEOS 2 observations are used together with the Earth's gravity field model GGM05S.

Although we are quite pleased with the analysis to date of frame dragging including LARES, LAGEOS and LAGEOS 2, we consider this result only intermediate to a final determination. Our final result will present a careful restudy of systematics.  We have been conservative here in quoting a 5\% estimate of our systematic error. Extending the observation time of LARES and the other satellites will improve our understanding of tidal contributions and will reduce the systematic error from that source. Different Earth's models lead to slightly different results, as is also the case for different orbital solvers. Completing a suite of solutions with different (up to date) Earth's models and different solvers will provide another estimate of the systematics. All these questions will be addressed in a forthcoming analysis of the measurement of frame-dragging using LARES, LAGEOS, LAGEOS 2 and GRACE.

However, we must also point out that the satellites LAGEOS, LAGEOS 2, and LARES will have tens of thousands of years on orbit, and will remain useful to laser-ranged science for an extremely long time. Eventually the retroreflectors may become degraded, but LAGEOS has shown no sign of this in its 40 years on orbit. Other laser-ranged satellites will be launched to join the current ones. All these satellites will be available while at the same time better Earth's models, better orbital solvers,  and better models of nongravitational forces become available. The strength of this approach and these satellites is that they are available for innovative improvements in technique into the future.

\section{Conclusions} Using the laser-ranged satellites LARES, LAGEOS and LAGEOS 2, and the Earth's gravity field described by the GGM05S model using the GRACE observations, we obtained a test of frame-dragging: $\mu = (0.994 \pm 0.002) \pm 0.05$, where $\mu = 1$ is the theoretical prediction of General Relativity, 0.002 is the 1-sigma statistical error and 0.05 is the estimated systematic error due to the uncertainties in the Earth's gravity field model GGM05S and to the other error sources.

\section{Acknowledgements}

We gratefully acknowledge the support of the Italian Space Agency, grants  I/034/12/0, I/034/12/1 and 2015-021-R.0 and
the International Laser Ranging Service for providing high-quality laser ranging tracking of the LARES satellites.
E.C. Pavlis acknowledges the support of NASA Grants NNX09AU86G and NNX14AN50G. R. Matzner acknowledges NASA  Grant NNX09AU86G and
J.C. Ries the support of NASA Contract NNG06DA07C.

%
%
%
%

\end{document}